\documentclass{osa-article}

\journal{oe}

\articletype{Research Article}
\usepackage{multirow}
\usepackage{makecell}
\begin{document}

\title{Practical quantum access network over a 10 Gbit/s Ethernet passive optical network}
\author{Bi-Xiao Wang,\authormark{1,2,6} Shi-Biao Tang,\authormark{3,6} Yingqiu Mao,\authormark{1,2} Wenhua Xu,\authormark{4} Ming Cheng,\authormark{5} Jun Zhang,\authormark{1,2} Teng-Yun Chen\authormark{1,2,7} and Jian-Wei Pan\authormark{1,2,8}}
\address{\authormark{1}Hefei National Laboratory for Physical Sciences at the Microscale and Department of Modern Physics, University of Science and Technology of China, Hefei 230026, China\\
\authormark{2}CAS Center for Excellence in Quantum Information and Quantum Physics, University of Science and Technology of China, Hefei 230026, China\\
\authormark{3}QuantumCTek Co., Ltd, Hefei 230088, China\\
\authormark{4}China Telecom Corporation Ltd. Shanghai Branch, Shanghai 200120, China\\
\authormark{5}Research Institute of China Telecom Co., Ltd, Shanghai 200122, China}

\email{\authormark{6}These authors contributed equally to this work.}
\email{\authormark{7}tychen@ustc.edu.cn} 
\email{\authormark{8}pan@ustc.edu.cn} 



\begin{abstract}
Quantum key distribution (QKD) provides an information-theoretically secure method to share keys between legitimate users. To achieve large-scale deployment of QKD, it should be easily scalable and cost-effective. The infrastructure construction of quantum access network (QAN) expands network capacity and the integration between QKD and classical optical communications reduces the cost of channel. Here, we present a practical downstream QAN over a 10 Gbit/s Ethernet passive optical network (10G-EPON), which can support up to 64 users. In the full coexistence scheme using the single feeder fiber structure, the co-propagation of QAN and 10G-EPON signals with 9 dB attenuation is achieved over 21 km fiber, and the secure key rate for each of 16 users reaches 1.5 kbps. In the partial coexistence scheme using the dual feeder fiber structure, the combination of QAN and full-power 10G-EPON signals is achieved over 11 km with a network capacity of 64-user. The practical QAN over the 10G-EPON in our work implements an important step towards the achievement of large-scale QKD infrastructure.
\end{abstract}

\section{Introduction}
Quantum key distribution (QKD) provides an effective solution to realize the remote secure communication by the laws of quantum mechanics\cite{Bennett1984,Gisin2002}. So far, remarkable progresses on QKD technology have been achieved, and various large-scale deployments of QKD have been reported\cite{Zhang:18}. Quantum access network (QAN) with point-to-multipoint connections is a practical approach to provide secure keys for multi-users and to extend the scale of QKD\cite{Townsend1997}. Meanwhile, the coexistence of QKD and classical optical communications in existing fiber infrastructures is a viable method to reduce the cost of the quantum channel and to improve the scalability of the QKD network\cite{Xu2020}. 

The classical access network is an important type of telecommunications network which connects end-users to network infrastructure. One of the most typical forms of classical access networks is the passive optical network (PON), where the downstream signal from the optical line terminal (OLT) at central office is broadcasted to all users by a power splitter, and the upstream signal from only one optical network unit (ONU) at user node is transmitted to OLT at each time slot\cite{Keiser2003}. Similar structure can also be used to build a QAN, which has two configurations, i.e., downstream (QKD receiver at user nodes) \cite{Townsend1997} and upstream (QKD transmitter at user nodes)\cite{Froehlich2015}. Therefore, realizing QAN over a PON provides secure keys for the "last mile" and greatly reduces the cost of optical fiber resources. 

The main challenge in the coexistence of QAN and PON is the spontaneous Raman scattering (SRS) noise generated by PON data signals\cite{Wang2015}. The effects of SRS noise are different for the upstream and downstream configurations. Due to backwards SRS, the upstream configuration suffers from higher noise\cite{Choi:10}. In addition, the implementation of the upstream QAN is more complicated, where one QKD receiver is allocated to all QKD transmitters by the time division multiplexing technology. One method is that different QKD transmitters alternately emit signal pulses to share the detector bandwidth. Yet, as the number of users increases, the difficulty in accurately assigning time slots will increase, and the secure key rate of each user will decrease significantly\cite{Froehlich2013}. The other method is that only one QKD transmitter occupies the detector at each time. Yet, after switching between different QKD transmitter, the QKD system link needs to be re-established, which increases the overall session time. The advantage of the upstream QAN is cost-effective, since the QKD receiver with single-photon detectors is relatively expensive. Compared with the upstream configuration, the downstream configuration has less SRS noise, and the secure key generation of each user is not influenced by other users, since each user holds SPDs independently from other users.

Many efforts have been taken to achieve the coexistence of QKD and PON. In the time-assignment scheme, the time-division multiplexing technique has been presented to reduce SRS noise, but it requires modification of the classical optical communication system\cite{Choi_2011}. In addition, other methods such as the post-processing scheme\cite{MartinezMateo2014} and the bypass structure scheme\cite{Sun2018} have been proposed to improve the signal-to-noise ratio of QKD. However, these schemes cannot realize operation for multi-user QANs. Recent work has demonstrated that the space-division multiplexing fiber provides additional isolation between quantum and classical signals, but such type of fiber is still far from practical applications\cite{Cai2020}. 

In this paper, we demonstrate a practical QAN, as well as its coexistence with a 10 Gbit/s Ethernet passive optical network (10G-EPON). Considering the fiber resources and power budget of the realistic 10G-EPON network, different coexistence schemes are proposed and experimentally implemented. In the full coexistence scheme based on the single feeder fiber structure, the co-propagation distance reaches 21 km with a secure key rate of 1.5 kbps for each of 16 users, when an additional attenuation of 9 dB is added to the 10G-EPON signal. Further, using the dual feeder fiber structure\cite{Choi:10,Froehlich2015,Vokic2020}, two partial coexistence schemes, i.e., dual feeder fiber coexistence and dual power splitter coexistence, are realized. In the dual feeder fiber coexistence, QAN integrated with full-power 10G-EPON signals supports up to 64 users with a secure transmission distance of 11 km. In the dual power splitter coexistence, the secure key rate of each user is improved and flexibly distributed by an independent quantum power splitter of QAN.

\section{Experiment}
We set up the coexisting access network based on the structure of the PON. In our experiment, a commercial 10G-EPON instrument (FiberHome AN5516-01) is used including one OLT, one optical distribution network (ODN), and three ONUs. For the QAN, QKD transmitter is placed with the OLT to form the central office, and each ONU is paired with a QKD receiver to form the user node. Details on the 10G-EPON pulses and QKD pulses are listed in Table 1. The ODN consists of three parts, i.e., the feeder fiber, the power splitter, and the drop fiber\cite{Tanaka2010}. The feeder fiber connects the central office with the power splitter, and the drop fiber connects the power splitter with the user nodes. In field environments, there are usually one or two feeder fiber links, while the number of drop fibers depends on the number of users. Typically, the distance of the drop fiber is shorter than that of the feeder fiber in order to save fiber resources. In this experiment, the distance of each drop fiber is fixed at 1 km. The single-mode fiber is based on ITU-T G.652.D and the fiber losses (Att) are also listed in Table 1. We note that the losses are higher than typical laboratory values to mimic field environments. In the 10G-EPON system, there is only one OLT laser. The 10G-OLT and 1G-OLT signals are integrated in an XFP transceiver (10 gigabit small form-factor pluggable), the average power of the overall OLT signal is 7.2 dBm. The number of ONU lasers equals to the number of users, which is three in this work. 10G-ONU, 1G-ONU, and 1G-ONU is assigned to User-1, User-2, and User-3, with peak powers of 5.7 dBm, 2.0 dBm, and 3.4 dBm, respectively. During the operation of quantum access network and the measurement of Raman noise, all four lasers are turned on. 

\begin{table}[htbp]
\renewcommand\arraystretch{2.38}
\centering
\caption{Description for the 10G-EPON and QKD pulses.}
\begin{tabular}{c|c|c|c|c|c}
\hline
 & \makecell{Wavelength range \\ (nm)} & \makecell{Center wavelength\\ (nm)}&\makecell{Data rate\\ (Gbps)}&Direction&\makecell{Att\\ (dB/km)}\\
\hline
1G-OLT& 1480$\sim$1500&1490&1.25&downstream&\multirow{2}*{0.31}\\

10G-OLT& 1575$\sim$1580&1577&10.3125&downstream&\\
\hline
10G-ONU&1260$\sim$1280&1270&10.3125&upstream&0.57\\

1G-ONU&1260$\sim$1360&1310&1.25&upstream&0.48\\

1G-ONU&1260$\sim$1360&1310&1.25&upstream&0.48\\
\hline
QKD-Sig&-&1550.12&-&downstream&0.35\\

QKD-Syn&-&1569.59&-&downstream&0.34\\
\hline
\end{tabular}
  \label{tab:shape-functions}
\end{table}

\begin{figure}[h!]
\centering\includegraphics[width=\textwidth]{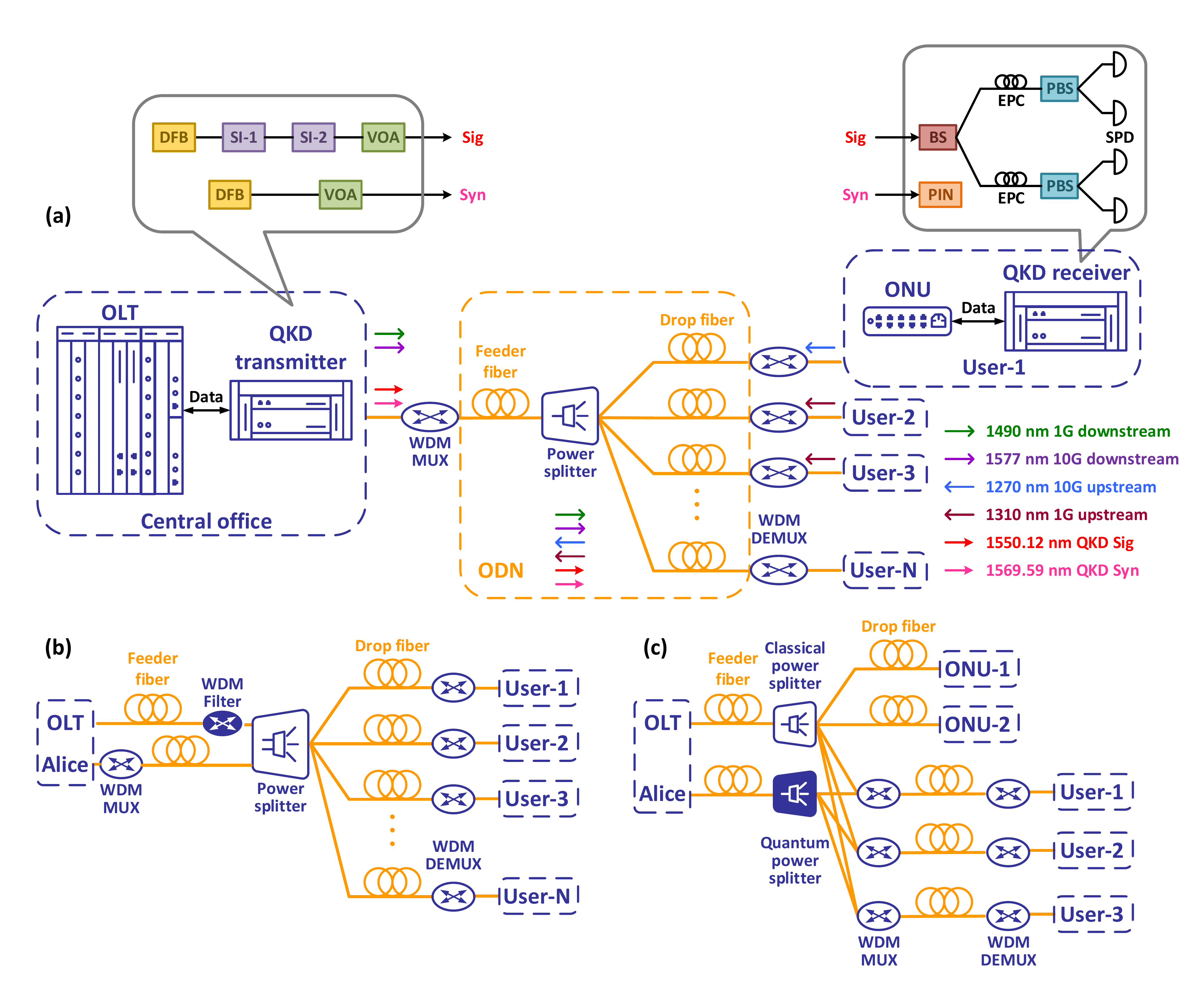}
\caption{Schematic layout of three coexistience schemes. (a) Full coexistence. (b) Dual feeder fiber scheme. (c) Dual power splitter scheme. WDM MUX/DEMUX: wavelength division multiplexer/demultiplexer. DFB: distributed feedback laser, SI: Sagnac interferometer, VOA: variable optical attenuator, BS: beam splitter, PBS: polarization beam splitter, EPC: electric polarization controller, SPD: single-photon detector, PIN: PIN photodetector. User-1, User-2, and User-3 are real users. The remaining users represent the virtual users.}
\end{figure}

The downstream QAN based on polarization-encoding decoy-state BB84 protocol\cite{Hwang2003,Wang2005,Lo2005}, includes one QKD transmitter and three QKD receivers, as illustrated in Fig. 1a. In the QKD transmitter, one distributed feedback laser is used to generate quantum optical pulses (Sig) with a repetition rate of 625 MHz\cite{Wang:20}. Each photon emitted by the QKD transmitter randomly chooses one output path of power splitter and is distributed to one of the QKD receivers. The different intensities and polarization states are implemented by two Sagnac interferometers in sequence\cite{Shen2013}. After attenuation, the average photon numbers of signal, decoy state, and vacuum state are 0.4, 0.1, and 0, with an emission ratio of 6:1:1, respectively. Also, the synchronized clock pulses (Syn) from QKD transmitter is broadcasted to all QKD receivers with a repetition rate of 100 kHz. In each QKD receiver, four InGaAs/InP SPDs are applied to measure the Sig pulses\cite{Liang2012,Zhang2015}. Each SPD operates at a gate frequency of 1.25 GHz, with a detection efficiency of $\sim$15\%, and a dark count rate per gate of $\sim$2$\times$$10^{-7}$.

For classical reconciliation, QKD transmitter is equipped with three modules, and each one corresponds to a receiver module at the three QKD receivers. The different classical reconciliation modules in QKD transmitter do not exchange data with each other.  In particular, the data traffic generated in the classical reconciliation stage is fed into the OLT and ONU to be exchanged through the network as 10G-EPON pulses. 

The calibration of the QAN includes delay scanning, synchronization calibration, and polarization feedback\cite{Chen2021}, and the specific optical signals emitted by the QKD transmitter is broadcasted to each QKD receiver in each procedure, thus the calibration is performed between QKD transmitter and each QKD receiver simultaneously. Key generation will only begin after all three links are successfully calibrated. 

The initial authentication of the public discussion channel between QKD transmitter and each QKD receiver is realized through the pre-stored keys, then periodically refreshed using the generated secure keys\cite{Mao:18}. In the post-processing, the Winnow algorithm\cite{Buttler2003} is used for error correction and the method of Toeplitz matrix\cite{Krawczyk1995} is applied for privacy amplification. Once the quantum bit error rate (QBER) exceeds the preset threshold of 4\%, QKD is aborted and the calibration of the QAN will restart. 

 \begin{figure}[h!]
\centering\includegraphics[width=\textwidth]{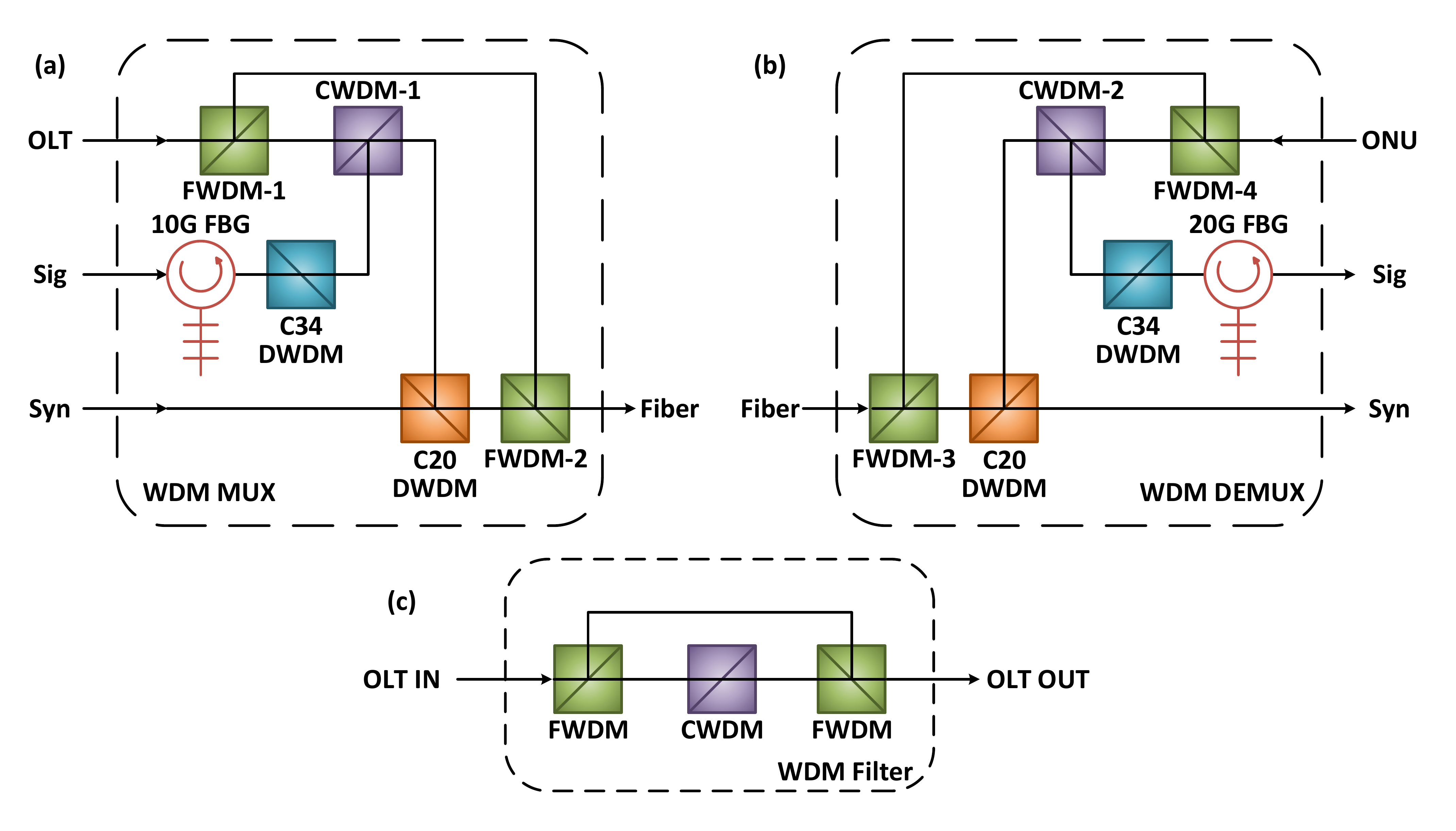}
\caption{The layout of WDM modules. (a) WDM MUX. (b) WDM DEMUX. (c) WDM Filter. FWDM: filter wavelength division multiplexer, passband 1510$\sim$1580 nm, reflectband 1260$\sim$1500 nm. CWDM: coarse wavelength division multiplexer, center wavelength 1578 nm. DWDM: 100 GHz dense wavelength division multiplexer, center wavelength 1550.12 nm (C34) and 1569.59 nm (C20). FBG: fiber Bragg grating, full width at half maximum 0.08 nm (10 GHz) and 0.16 nm (20 GHz).} 
\end{figure}

In the single feeder fiber structure, the full coexistence of the QAN and 10G-EPON over the total ODN is achieved, as depicted in Fig. 1a. The coexistence is realized using a WDM MUX and DEMUX module, as illustrated in Fig. 2a and 2b. In the WDM MUX module, the isolation of 1550.12 nm from 1G-OLT signal is 23 dB by FWDM-1, and then the band-pass filter of 1G-OLT signal is realized by FWDM-2. Using the CWDM-1, the band-pass filter of 10G-OLT signal and the band-stop filter of 1550.12 nm component in 10G-OLT signal are achieved. In the WDM DEMUX module, both the isolation of ONU signal and 1G-OLT signal from 1550.12 nm are 107 dB, and the isolation of 10G-OLT signal from 1550.12 nm is 71 dB. The WDM modules provide sufficient cross-talk isolation, and the linear crosstalk of 10G-EPON signals is eliminated effectively. 

The insertion losses of OLT, 10G-ONU, 1G-ONU and quantum signals in WDM MUX/DEMUX are 0.9/1.0 dB, 1.0/0.5 dB, 0.7/0.5 dB, and 0.8/3.4 dB, respectively. The network capacity of the coexistence network corresponds to the splitting ratio of power splitter, i.e., 1:4, 1:8, 1:16, 1:32, and 1:64, where their average insertion losses are 7.4 dB, 10.5 dB, 13.6 dB, 17.1 dB, and 20.2 dB, respectively.

Further, we investigate the dual feeder fiber structure. The second feeder fiber is assigned as quantum channel, which is generally a spare fiber for the PON. Two possible partial coexistence schemes, i.e., dual feeder fiber scheme and dual power splitter scheme, are depicted in Fig. 1b and 1c. In the partial coexistence schemes, the linear crosstalk of the OLT laser and the nonlinear SRS noise caused by the OLT signal in the feeder fiber are filtered with a WDM Filter module, before the quantum and classical signals are combined by the power splitter, effectively lowering the noise photons, as shown in Fig. 2c. In flexible network scenarios where the users can freely choose to be connected or not to the QAN, the dual power splitter scheme can be applied, where the quantum signals travel through a separate quantum power splitter from the classical signals.

\section{Results and discussion}

The SRS noise photons in the proposed coexistence network schemes are mainly generated by the OLT signal. The SRS noise caused by the ONU signal can be negligible, since the wavelength interval between ONU and quantum signals is more than 190 nm apart\cite{Wang2017}. The OLT signal generates SRS noise photons in both the feeder fiber and drop fiber, which is described as\cite{Chapuran_2009,Sun2018}, 
\begin{equation}
S_F=P\frac{ \beta}{\alpha_{q}-\alpha_{c}} (e^{-\alpha_{c} L_F}-e^{-\alpha_{q} L_F})\frac{1}{N}e^{-\alpha_{q}L_D},
\end{equation}

\begin{equation}
S_D=\frac{P}{N} e^{-\alpha_{c}L_F}\frac{\beta}{\alpha_{q}-\alpha_{c}} (e^{-\alpha_{c} L_D}-e^{-\alpha_{q} L_D}),
\end{equation}
where $\alpha_{c}$ and $\alpha_{q}$ correspond to the path attenuation coefficient of OLT and quantum signal, $L_F$ and $L_D$ are the distance of feeder fiber and drop fiber, $P$ is the launch power of the OLT signal, $N$ is the splitting ratio of power splitter, and $\beta$ is the SRS coefficient.

In the single feeder fiber structure, i.e., full coexistence, the total SRS noise detectable by QKD receiver is the sum of $S_F$ and $S_D$. The SRS noise in the full coexistence scheme is affected by three factors, i.e., the splitting ratio of power splitter, the distance of feeder fiber, and the launch power of the OLT signal. 

\begin{figure}[h!]
\centering\includegraphics[width=\textwidth]{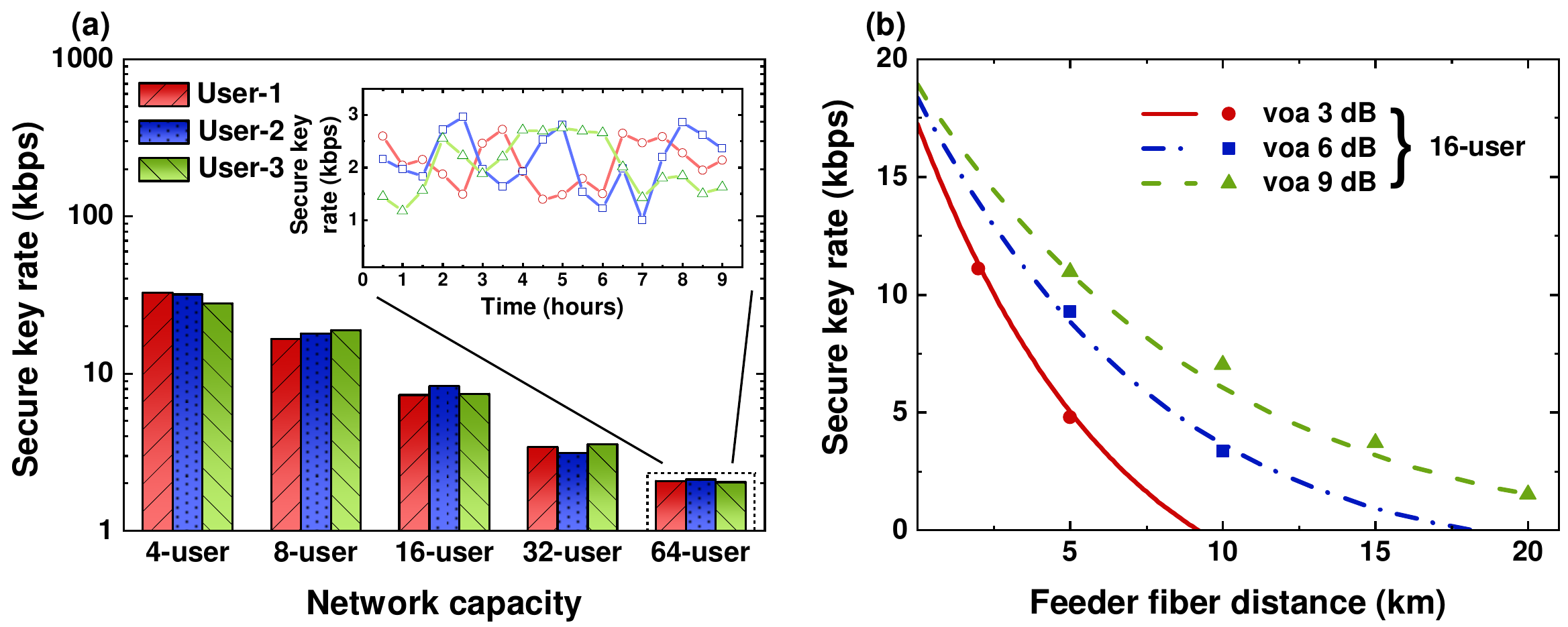}
\caption{Measured and simulated results for full coexistence scheme. (a) Secure key rate of each user under different network capacity. The distance of feeder fiber is fixed at 5 km. An attenuation of 5 dB is added to the OLT signal. Inset: secure key rate of each user verses time in a 64-user network. (b) User-1 secure key rate as a function of feeder fiber distance under different attenuation of the OLT signal. The splitting ratio is fixed at 1:16. The red solid, blue dash-dotted, and green dashed line represents the simulation under the 3 dB, 6 dB, and 9 dB attenuation of the OLT signal, respectively. Red circles, blue squares, and green triangles denote the experimental data under the 3 dB, 6 dB, and 9 dB attenuation of the OLT signal, respectively.}
\end{figure}

First, the splitting ratio of power splitter is adjusted, while the distance of feeder fiber and the launch power of the OLT signal are fixed. The results are shown in Fig. 3a. In addition, a key rate stability test of the QAN when the network capacity is 64 users is shown in the inset of Fig. 3a. The total transmission loss from QKD transmitter to QKD receiver is $\sim$26.5 dB, and an average key rate of 2.1 kbps per user is obtained over 9 hours.

Second, to investigate the maximum secure transmission distance of the QAN under different launch powers of the OLT signal, the splitting ratio of power splitter is fixed. The measured and simulated secure key rates of each user under different feeder fiber distance and the launch power of the OLT signal are plotted in Fig. 3b. The simulation is performed with methods from Ref.\cite{Ma2005}, and the secure key rate, $R$, is given by 
\begin{equation}
R=q\lbrace Q_1[1-H_2(e_1)]-fQ_{\mu}H_2(E_{\mu})+Q_0\rbrace,
\end{equation}
where $q$ is the basis-sift factor, which is 1/2 in this experiment, $f$ is the correction efficiency of error correction, which is 1.2$\sim$1.5, $H_2(x)$ is the binary entropy function, $Q_\mu$ and $E_\mu$ are the overall gain and QBER of the signal state, $Q_1$ and $e_1$ are the gain and QBER of the single-photon of signal states, $Q_0$ is the background gain, which comes from the sum of dark and Raman noise ($S_F$ and $S_D$) counts in the full coexistence scheme.

We note that without additional attenuation to the OLT signal, the distance of feeder fiber only reaches 1 km. As the distance increases, the launch power of the OLT signal should be decreased. In order to generate secure keys over a typical distance for PON as long as 20 km of feeder fiber, the OLT signal must be attenuated by 9 dB. Nevertheless, the additional attenuation on the classical signals may affect the performance of the 10G-EPON, which requires its pulse power to be within the power budget.

\begin{table}[h!]
\centering
\caption{Comparison of SRS noise photon count rate (kcps) in the full and partial coexistence schemes. }
\begin{tabular}{ccc}
\hline
Feeder fiber distance & Full coexistence & Partial coexistence \\
\hline
5 km & 16.3 & 2.9 \\
20 km & 18.1 & 1.0\\
\hline
\end{tabular}
  \label{tab:shape-functions}
\end{table}

Further, we investigate the partial coexistence scheme. In such scheme, $S_F$ is eliminated before the multiplexing of the quantum and OLT signals, therefore, the SRS noise is only contributed by $S_D$. The measured SRS noise generated by the full-power OLT signal in the 16-user network of the full and partial coexistence schemes are listed in Table 2, from which one can see the SRS noise photons are drastically lowered with the partial coexistence scheme. Therefore, the QAN can be combined with full-power 10G-EPON over long distance ODN in the partial coexistence scheme.

\begin{figure}[h!]
\centering\includegraphics[width=\textwidth]{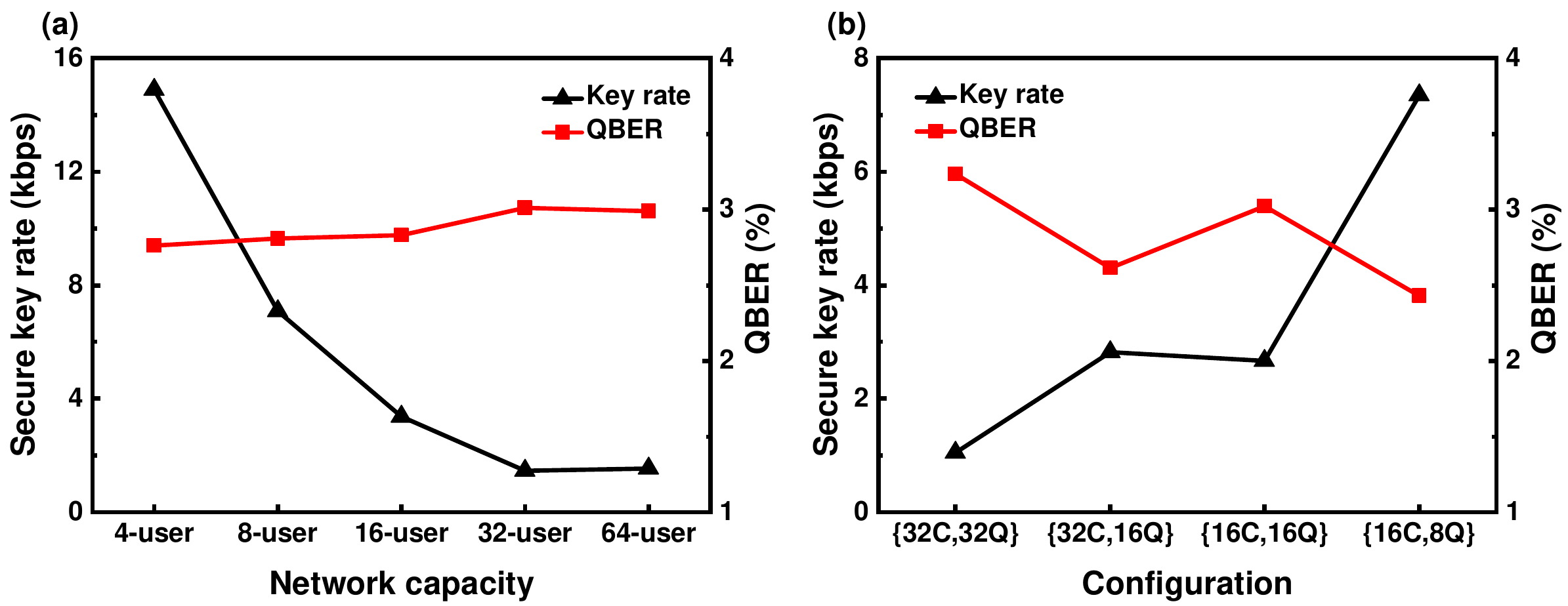}
\caption{Results for the partial coexistence schemes. Black triangles and red squares denote the experimental data of secure key rate and QBER. No attenuation is added to the OLT signal. (a) User-1 secure key rate and QBER under different network capacity in dual feeder fiber scheme. For the 4-user, 8-user, 16-user, and 32-user networks, the distance of feeder fiber is 20 km. In the 64-user network, the distance of feeder fiber is 10 km. (b) User-1 secure key rate and QBER under different configurations in the dual power splitter scheme. In each configuration, the first and the second are the network capacities of classical and quantum network, respectively. The distance of the feeder fiber is 20 km.}
\end{figure}

Figure 4a shows the performance of secure key rate and QBER in the dual feeder fiber scheme. When the total distance of ODN is fixed at 21 km, the maximum network capacity is 32 users. To support network capacity of 64 users, the maximum secure transmission distance of ODN is 11 km.

In the dual power splitter scheme, the network capacities of 10G-EPON and QAN are independent. The performance of the QAN under four network capacity configurations is shown in Fig. 4b. The secure key rate is increased by using a power splitter with lower splitting ratio for the quantum signals, see the configuration between $\lbrace$32C,32Q$\rbrace$ and $\lbrace$32C,16Q$\rbrace$, and $\lbrace$16C,16Q$\rbrace$ and $\lbrace$16C,8Q$\rbrace$. Compared with the $\lbrace$32C,16Q$\rbrace$ configuration, the splitting ratio of the classical power splitter is lower in the $\lbrace$16C,16Q$\rbrace$ configuration, and thus, the launch power of the OLT signal and the Raman noise generated by OLT signal in the drop fiber is higher, and the secure key rate is lower in the $\lbrace$16C,16Q$\rbrace$ configuration. Furthermore, the dual power splitter coexistence scheme has an additional benefit that it is highly flexible. The splitting ratio of power splitter can be imbalance, which is customized according to the user demand in this scheme, under the premise that the user with the minimum splitting ratio can also obtain the secure key.

\section{Conclusion}
In summary, we have demonstrated a practical downstream QAN over a 10G-EPON with the maximum network capacity of 64-user. Based on the optical fiber topology and power budget of 10G-EPON, different coexistence schemes, i.e., full coexistence, dual feeder fiber coexistence, and dual power splitter coexistence, are proposed, which are suitable for different field network environments. The full coexistence scheme based on the single feeder fiber structure is suitable for scenarios where the 10G-EPON has sufficient power budget. When the 10G-EPON signal is attenuated by 9 dB, QAN supports 16 users with a secure key rate of 1.5 kbps per user and the secure transmission distance reaches 21 km. In the partial coexistence schemes, the SRS noise from feeder fiber is effectively eliminated by the dual feeder fiber structure, and QAN is integrated with full-power 10G-EPON signals. The secure key rate of each user can be customized according to the user demand in the dual power splitter coexistence scheme. Our work provides a practical and flexible method to implement the coexistence of QAN and 10G-EPON, which could be favourable for the deployment of QKD network infrastructure.

\begin{backmatter}
\bmsection{Funding}
China State Railway Group Co., Ltd. Scientific and Technological Research Project (K2019G062); National Natural Science Foundation of China (61875182); Anhui Initiative in Quantum Information Technologies.

\bmsection{Disclosures}
The authors declare no conflicts of interest.

\bmsection{Data availability} Data underlying the results presented in this paper are not publicly available at this time but may be obtained from the authors upon reasonable request.
\end{backmatter}

\bibliography{sample}

\end{document}